\documentclass[twocolumn]{{aastex7}}

\begin{document}

\title{Pre-nova Observations of T CrB: A view from the CHARA Array}

\correspondingauthor{Ryan Norris}
\email{ryan.norris@nmt.edu}

\author[0000-0002-9120-9728]{Ryan Norris}
\affiliation{New Mexico Institute of Mining and Technology, Workman Center, 801 Leroy Place, Socorro, NM 87801, USA}
\email{ryan.norris@nmt.edu}
\author[0000-0002-2208-6541]{Narsireddy Anugu}
\affiliation{The CHARA Array of Georgia State University, Mount Wilson Observatory, Mount Wilson, CA 91023, USA}
\email{nanugu@gsu.edu}

\author[0009-0000-6957-8466]{Thomas Gaudin}
\affiliation{ Department of Astronomy and Astrophysics, The Pennsylvania State University, University Park, PA 16802, USA}
\email{tmg6006@psu.edu}

\author[0000-0003-1564-7029]{Magdalena Otulakowska-Hypka}
\affiliation{Astronomical Observatory Institute, Faculty of Physics and Astronomy, Adam Mickiewicz University, S\l{}oneczna 36, PL-60286 Poznań, Poland}
\email{magdaot@amu.edu.pl}

\author[0000-0002-3445-919X]{Fabian Kaczmarek}
\affiliation{Astronomical Observatory Institute, Faculty of Physics and Astronomy, Adam Mickiewicz University, S\l{}oneczna 36, PL-60286 Poznań, Poland}
\email{fabkac@amu.edu.pl}

\author{Cameron Caruso}
\affiliation{New Mexico Institute of Mining and Technology, Workman Center, 801 Leroy Place, Socorro, NM 87801, USA}
\email{cameron.caruso@student.nmt.edu}

\author{Cody Gustafson}
\affiliation{New Mexico Institute of Mining and Technology, Workman Center, 801 Leroy Place, Socorro, NM 87801, USA}
\email{cody.gustafson@student.nmt.edu}

\author{Andrew Kotowski}
\affiliation{New Mexico Institute of Mining and Technology, Workman Center, 801 Leroy Place, Socorro, NM 87801, USA}
\email{andrew.kotowski@student.nmt.edu}

\author{Rebecca Proni}
\affiliation{New Mexico Institute of Mining and Technology, Workman Center, 801 Leroy Place, Socorro, NM 87801, USA}
\email{rebecca.proni@student.nmt.edu}

\author{Nirupam Roy}
\affiliation{New Mexico Institute of Mining and Technology, Workman Center, 801 Leroy Place, Socorro, NM 87801, USA}
\email{nirupam.roy@student.nmt.edu}

\author[0000-0002-8376-8941]{Fabien Baron}
\affiliation{Center for High Angular Resolution Astronomy and Department 
of Physics and Astronomy, Georgia State University, P.O. Box 5060, Atlanta,
GA 30302-5060, USA}
\email{fbaron@gsu.edu}

\author[0000-0003-4896-2543]{Dipankar P.K Banerjee}
\affiliation{Astronomy and Astrophysics Division, Physical Research Laboratory, Ahmedabad, India 380009}
\email{dpkb12345@gmail.com}

\author[0000-0002-9512-9792]{Dana K. Baylis-Aguirre}
\affiliation{New Mexico Institute of Mining and Technology, Workman Center, 801 Leroy Place, Socorro, NM 87801, USA}
\email{dana.baylisaguirre@nmt.edu}

\author[0000-0002-8349-9366]{Michelle J. Creech-Eakman}
\affiliation{New Mexico Institute of Mining and Technology, Workman Center, 801 Leroy Place, Socorro, NM 87801, USA}
\email{Michelle.CreechEakman@nmt.edu}

\author[0000-0002-3873-5497]{Justin Linford}
\affiliation{National Radio Astronomy Observatory, Domenici Science Operations
Center, 1011 Lopezville Road, Socorro, NM 87801, USA}
\email{jlinford@nrao.edu}

\author[0000-0001-7853-4094]{Alexandre Gallenne}
\affiliation{Instituto de Alta Investigaci\'on, Universidad de Tarapac\'a, Casilla 7D, Arica, Chile}
\email{agallenne@academicos.uta.cl}

\author[0000-0003-3457-0020]{ Joanna Miko{\l}ajewska}
\affiliation{Nicolaus Copernicus Astronomical Center, Polish Academy of Sciences, Bartycka 18, PL-00-716 Warsaw, Poland}
\email{mikolaj@camk.edu.pl}

\author[0000-0002-3380-3307]{John D. Monnier}
\affiliation{Department of Astronomy, University of Michigan, Ann Arbor, MI 48109, USA}
\email{monnier@umich.edu}

\author[0000-0001-7425-5055]{Denis Mourard}
\affiliation{Universit\'e C\^ote d'Azur, Observatoire de la C\^ote d'Azur, CNRS, Laboratoire Lagrange, France}
\email{denis.mourard@oca.eu}

\author[0000-0001-6805-9664]{Ulisse Munari}
\affiliation{INAF National Institute of Astrophysics, Astronomical Observatory of Padova, 36012 Asiago (VI), Italy }
\email{ulisse.munari@inaf.it}

\author[0000-0002-7399-0231]{Nicolas Nardetto}
\affiliation{Laboratoire LAGRANGE - UMR 7293
UNS / CNRS / OCA}
\email{Nicolas.Nardetto@oca.eu}

\author[0000-0002-9288-3482]{Rachael M.\ Roettenbacher}
\affiliation{Department of Astronomy, University of Michigan, Ann Arbor, MI 48109, USA}
\email{rmroett@umich.edu}

\author[0000-0003-2835-0304]{Jennifer L Sokoloski}
\affiliation{Columbia Astrophysics Laboratory and Department of Physics, Columbia University, New York, NY 10027, USA}
\email{jeno@astro.columbia.edu}

\author[0009-0006-3000-0900]{Montana Williams}
\affiliation{New Mexico Institute of Mining and Technology, Workman Center, 801 Leroy Place, Socorro, NM 87801, USA}
\email{montana.williams@student.nmt.edu}

\author[0000-0001-6567-627X,sname='Woodward']{C. E. Woodward}
\affiliation{Minnesota Institute for Astrophysics, School of Physics \& Astronomy, 116 Church Street SE, University of Minnesota, Minneapolis, MN 55455, USA}
\email{chickw024@gmail.com}

\author[0000-0001-6017-8773]{Stefan Kraus}
\affiliation{Astrophysics Group, Department of Physics \& Astronomy, University of Exeter, Stocker Road, Exeter, EX4 4QL, UK}
\email{S.Kraus@exeter.ac.uk}

\author[0000-0002-0493-4674]{Jean-Baptiste Le Bouquin}
\affiliation{Institut de Plan\'etologie et d'Astrophysique de Grenoble, 38058 Grenoble, France}
\email{jean-baptiste.lebouquin@obs.ujf-grenoble.fr}

\author[0000-0001-8926-9732]{Sorabh Chhabra}
\affiliation{Astrolhysics Group, Department of Physics \& Astronomy, University of Exeter, Stocker Road, Exeter, EX4 4QL, UK}
\email{S.Chhabra@exeter.ac.uk}

\author[0009-0005-8088-0718]{Isabelle Codron}
\affiliation{Astrophysics Group, Department of Physics \& Astronomy, University of Exeter, Stocker Road, Exeter, EX4 4QL, UK}
\email{ic302@exeter.ac.uk}

\author[0000-0002-1575-4310]{Jacob Ennis}
\affiliation{Department of Astronomy, University of Michigan, Ann Arbor, MI 48109, USA}
\email{ennisj@umich.edu}

\author[0000-0002-3003-3183]{Tyler Gardner}
\affiliation{Astrophysics Group, Department of Physics \& Astronomy, University of Exeter, Stocker Road, Exeter, EX4 4QL, UK}
\email{tgardne@umich.edu}

\author[0009-0006-0225-4444]{Mayra Gutierrez}
\affiliation{Department of Astronomy, University of Michigan, Ann Arbor, MI 48109, USA}
\email{mgutie60@ucsc.edu}

\author[0000-0002-1788-9366]{Noura Ibrahim}
\affiliation{Department of Astronomy, University of Michigan, Ann Arbor, MI 48109, USA}
\email{inoura@umich.edu}

\author[0000-0001-9745-5834]{Cyprien Lanthermann}
\affiliation{The CHARA Array of Georgia State University, Mount Wilson Observatory, Mount Wilson, CA 91023, USA}
\email{cyprien@chara-array.org}

\author[0000-0001-5980-0246]{Benjamin R. Setterholm}
\affiliation{Max-Planck-Institut für Astronomie, Heidelberg, Germany}
\email{setterholm@mpia.de}

\author[0000-0001-9939-2830]{Christopher D. Farrington}
\affiliation{The CHARA Array of Georgia State University, Mount Wilson Observatory, Mount Wilson, CA 91023, USA}
\email{farrington@chara-array.org}

\author{Olli Majoinen}
\affil{The CHARA Array of Georgia State University, Mount Wilson Observatory, Mount Wilson, CA 91023, USA}
\email{olli@chara-array.org}

\author{Norm Vargas}
\affil{The CHARA Array of Georgia State University, Mount Wilson Observatory, Mount Wilson, CA 91023, USA}
\email{nvargas@gsu.edu}

\author[0000-0001-8537-3583]{Douglas R. Gies}
\affiliation{Center for High Angular Resolution Astronomy and Department 
of Physics and Astronomy, Georgia State University, P.O. Box 5060, Atlanta,
GA 30302-5060, USA}
\email{gies@chara.gsu.edu}

\author[0000-0001-5415-9189]{Gail H. Schaefer}
\affiliation{The CHARA Array of Georgia State University, Mount Wilson Observatory, Mount Wilson, CA 91023, USA}
\email{schaefer@chara-array.org}

\author[0000-0002-0114-7915]{Theo ten Brummelaar}
\affiliation{The CHARA Array of Georgia State University, Mount Wilson Observatory, Mount Wilson, CA 91023, USA}
\email{theo@chara-array.org}

\begin{abstract}
T CrB is a symbiotic recurrent nova consisting of a red giant and white dwarf with recent eruptions in 1866 and 1946 and an anticipated eruption in the mid 2020s. We report CHARA Array observations obtained with MIRC-X (\textit{H}-band) and MYSTIC (\textit{K}-band) in 2022--2025. We fit limb darkened disk models constrained with literature limb darkening coefficients to the squared visibilities as only the first visibility lobe is sampled. The average limb darkened diameter of the star across these epochs is $0.70\pm0.04$ mas in \textit{H}-band and $0.72\pm0.07$ mas in \textit{K}-band. Adopting a distance of $914^{+24}_{-22}$ pc, the stellar radius is $69\pm5~R_{\odot}$ in \textit{H}-band  and $71\pm8~R_{\odot}$ in \textit{K}-band. This is consistent with filling a Roche lobe volume radius of $71~R_{\odot}$ inferred from published orbital solutions. These measurements provide a pre-eruption angular diameter and support a Roche lobe filling donor.
\end{abstract}

\keywords{ Binary stars (154) --  High angular resolution (2167) -- Interferometric binary stars (806)   -- Long baseline interferometers (931) -- Optical interferometry (1168) -- Recurrent novae (1366) -- Symbiotic binary stars (1674)}

\section{Introduction} \label{sec:intro}
Symbiotic recurrent novae are a group of interacting binaries consisting of a red giant (RG) and white dwarf (WD). The description ``recurrent'' refers to occurrence of eruptions on a time scale of generally less than 100 years, whereas the name symbiotic identifies these as a particular group of interacting binaries that exhibit both absorption lines from a cool star and emission lines from gas excited by a hot companion. T Coronae Borealis (T CrB, HD 143454) is the closest symbiotic recurrent nova at a distance of $914^{+24}_{-22}~\textrm{pc}$ \citep{schaeferdistances}. It last erupted on February 9, 1946 and prior to that, experienced an eruption in 1866 \citep{schaefer_b_2023}, suggesting a recurrence time of 80 years or 128 orbits \citep{munari_secondary_2023}. Historical research suggests eruptions of the system were also recorded in 1217 and 1787 \citep{schaefer_recurrent_2023}. The proximity of the system and the expectation of an upcoming eruption is generating considerable interest in the astronomical community.

The behavior of the system during the past ten years has supported the expectation that another eruption will occur around the 80 year recurrence rate. \citet{munari_superactive2016} reported an increase in brightness, changes in the light curve, and increases in the strength and number of high ionization lines, all of which pointed to an increase in activity in the system. \citet{luna_increasing_2020} found similarities between light curves obtained prior to the 1946 eruption and those obtained after the start of the super-active state in 2014, suggesting that super-active states lead up to eruptions.  \citet{zamanov_pre-outburst_2023} used \textit{UBV} photometry to measure the mass accreted during the super-active state and supported \citet{luna_increasing_2020}'s theory that a significant portion of the mass required for a thermonuclear runaway is accreted during super-active states. By 2023, the system had dropped out of the super-active state \citep{weakeningother,swiftweak,halphaweak}. However, photometric and spectroscopic changes in 2024 indicated a return to a high accretion state \citep{2025MNRAS.541L..14M}. \citet{2026A&A...706A..94P} connected photometric dips in 2023-2024 \citep[e.g.,][]{dip} and 2024-2025 to soft X-ray brightening, which the authors attribute to a thinning of the boundary layer between the  WD and the inner accretion disk, noting that this could suggest that the WD is approaching thermonuclear runaway. Nonetheless, \citet{2025MNRAS.541L..14M} recommend caution when using the timing of previous eruptions and the phenomena preceding them to estimate the timing of an upcoming eruption, warning that the  exact timing remains unpredictable due to highly variable accretion rates. 

\citet{2025hinkle} reports that the angular diameter of the RG in T CrB was measured in the near-infrared \textit{H} and \textit{K}-bands using 18 nights of archival interferometric observations collected with the Palomar Testbed Interferometer (PTI) between 2003 November and 2005 February, resulting in an average uniform disk angular diameter of $0.85\pm0.15$ mas \citep{Rogge2011}. However, this angular diameter is beyond the resolving limit of PTI at these wavelengths given its longest baseline of 110 m \citep{Colavita_1999, 2008ApJS..176..276V}. Thus, additional measurements of the angular diameter of the RG in the system are needed. The angular diameter of the system, when used with a distance, yields a radius, which is important not only for comparison to the theoretical Roche-lobe radius, but also for analysis and modeling of the post-eruption ejecta. 

As of the time of submission, T CrB has not yet erupted. In anticipation of the upcoming eruption, we began an observing campaign with the Center for High Angular Resolution Astronomy (CHARA) Array with the goal of collecting optical interferometric observations prior to, during, and after the upcoming eruption. Here we report measurements of the angular diameter of the RG in the system, which will provide a baseline for comparison during eruption, as well as a test of different models of the system.

\section{Observations} \label{sec:obs}
\begin{deluxetable*}{l c c l c c}
\tablecaption{CHARA observations of T CrB. The calibrator star diameters adapted from the JMMC catalog \citep{Bourges2017}. \label{Table_obs_log}}
\tablewidth{0pt}
\tablehead{
    \colhead{DATE} & \colhead{MIRC-X} & \colhead{MYSTIC} & \colhead{Calibrators} & \colhead{$\theta_{\text{UD,H}}$} & \colhead{$\theta_{\text{UD,K}}$} \\
    \colhead{(UT)} & \colhead{($\mathcal{R}=\lambda / \triangle \lambda$)} & \colhead{($\mathcal{R}=\lambda / \triangle \lambda$)} & \colhead{} & \colhead{(mas)} & \colhead{(mas)}
}
\startdata      
2022-03-03 & 190 & --- & HD 142053 & $0.50\pm0.01$ & --- \\
2023-05-09 & 190 & 278 & HD 142053 & $0.50\pm0.01$ & $0.51\pm0.01$ \\
           &     &     & HD 149067 & $0.36\pm0.05$ & $0.37\pm0.05$ \\
2023-06-04 & 190 & 278 & HD 149067 & $0.36\pm0.05$ & $0.37\pm0.05$ \\
2023-09-09 & 190 & --- & HD 142053 & $0.50\pm0.01$ & --- \\
2024-06-22 & 50  & 278 & HD 134083 & $0.65\pm0.06$ & $0.66\pm0.06$ \\
           &     &     & HD 133485 & $0.65\pm0.05$ & $0.66\pm0.06$ \\
2024-06-23 & 50  & 49  & HD 134083 & $0.65\pm0.06$ & $0.66\pm0.06$ \\
           &     &     & HD 145457 & $0.67\pm0.05$ & $0.68\pm0.05$ \\
2025-02-26 & ---  & 49  &HD 143393 & --- & $0.64\pm0.06$ \\ 
\enddata
\end{deluxetable*}
We observed T CrB on 9 nights from 2022 to 2025 using the \textit{H}-band MIRC-X \citep[$\lambda=1.5 - 1.8 ~\mu$m,][]{Anugu2020} and the \textit{K}-band MYSTIC beam combiners \citep[$\lambda=2.0 - 2.4 ~\mu$m,][]{Setterholm2023} on the CHARA Array, located at the Mount Wilson Observatory, California, USA. The CHARA Array consists of six 1-meter telescopes arranged in a Y-shaped configuration \citep{tenBrummelaar2005}, making it the largest optical and near-infrared interferometer in the world. The array has baseline lengths ranging from 34 meters to 331 meters. Both MIRC-X and MYSTIC combine beams from all six telescopes, providing 15 squared visibilities ($V^2$) and 10 independent closure phases within each operational wavelength band. MIRC-X  covers 30 spectral channels in the \textit{H}-band in  the spectral resolution mode of $\mathcal{R}=190$, while MYSTIC spans 52 spectral channels in the \textit{K}-band in the spectral resolution mode of $\mathcal{R}=278$. These instruments offer angular resolutions of approximately 0.5 milliarcseconds (mas) in the \textit{H}-band and 0.7 mas in the \textit{K}-band. 

The dates of observations, instrument mode used, and calibrator details are presented in Table \ref{Table_obs_log}. On some nights, either MYSTIC or MIRC-X was unavailable for use and observations were only collected using one instrument. Observations collected on two nights in February 2025 did not result in useful data due to poor weather conditions and are not used in this study. In 2022 and 2023, observations were collected using higher resolution modes ($\mathcal{R}=190$ for MIRC-X, $\mathcal{R}=278$ for MYSTIC) because the observations took place during a larger program to study symbiotic stars. In 2024 and 2025, T CrB was observed as an additional target during unrelated observing campaigns and observations were collected at lower spectral resolutions. 

Due to practical observing conditions, such as instrumental fringe smearing during exposure, polarization mismatches, baseline-dependent factors, and atmospheric turbulence, the measured $V^2_{\rm star, measured}$ can be corrupted from the true value. To correct this, the data are calibrated ($V^2_{\rm star, calibrated}$) by comparing the measurements with the visibilities of a nearby reference (calibrator) star with a known visibility. We selected calibrators using \texttt{SearchCal} based on the following criteria: (i) nearly unresolved point sources in the \textit{H}-band to minimize biases due to the assumed calibrator star diameter, (ii) confirmed through binarity checks using from \textit{Gaia} Data Release 3 (DR3; \citealt{gaiadr3doi}), as described in \citet{2016A&A...595A...1G} and \citet{2023A&A...674A...1G}; (iii) bright calibrators with less than a 1-magnitude difference from the target to ensure a higher signal-to-noise ratio; and (iv) minimal angular separation from the target. Table~\ref{Table_obs_log} lists the calibrators, along with their uniform disk angular diameters and error bars ($\theta_{\rm UD}$) obtained from the JMMC catalog \citep{Bourges2017}.

When possible, we used at least two calibrators with less than 30 minutes separating observations of a calibrator and the target, collected on all 15 baselines, used a CAL-SCI-CAL-SCI bracket sequence, and observed on more than one night in order to minimize calibration systematics. At times, it was not possible to observe on multiple nights or obtain multiple brackets due to the constraints of weather or program schedules. On the nights of UT 2022-03-03, 2023-05-09, and 2024-06-22, the time gap between the calibrator observation and target observation was over 50 minutes, resulting in increased uncertainty in the angular diameters measured using data collected on these nights. The raw MIRC-X and MYSTIC data sets\footnote{available at \href{https://www.chara.gsu.edu/observers/database}{ https://www.chara.gsu.edu/observers/database}} were reduced using the standard MIRC-X~pipeline, version 1.4.0 \citep{Anugu2020,le_bouquin_2024_12735292}. 

For data collected in 2022 and 2023, we calibrated each calibrator against the other on nights when two calibrators were used (2023 May 09). Using updated calibrator diameters changed the measured angular diameters of T CrB by less than the uncertainty in the measurement. For analysis in this paper, we use data calibrated using the diameters from \citet{Bourges2017}, as listed in Table \ref{Table_obs_log}.  In addition, to test the impact of calibrators on the 2024-06-22 and 2024-06-23 observations, we calibrated observations for those nights three different ways: 1) using both the first and second calibrator for each night; 2) using only the first calibrator; 3) using only the second calibrator. We found that the different calibrator combinations resulted in angular diameters measurements of the RG within the average uncertainty of \textit{H}-band diameters, and slightly above than the average uncertainty of the \textit{K}-band diameters. The angular diameters reported in this paper are the results of fits to data calibrated using both calibrator stars on these nights. However, we caution that the angular diameters of the calibrator stars used in 2024 and 2025 are close to the angular diameter of the RG in T CrB, so the calibrated data on these nights is especially sensitive to any inaccuracy in the calibrator diameters used. 
 
The calibrated observations consist of $V^2$, triple amplitudes, and closure phases. Closure phases did not vary significantly from zero, with a maximum deviation from zero of no more than $1.56\pm0.14 ^{\circ}$ in \textit{H}-band and $2.14\pm0.26^{\circ}$ in \textit{K}-band across all epochs, indicating that no significant departure from centro-symmetry was detected at the spatial scales we could resolve.
\
\section{Results} \label{sec:analysis}
\subsection{Companion Search}

Although we do not expect to detect the WD or accretion disk in the system at near-infrared wavelengths prior to an eruption \citep{2012BASI...40..267C,maslennikova_recurrent_2023}, we used CANDID \citep{candid} to search for tertiary companions. Due to the large error on $V^{2}$, we used only the closure phases in the search for companions. Although some spurious high-$\sigma$ detections were recorded within a 10 mas radius, we deemed these to be related to artifacts due to their inconsistency across neighboring nights and lack of relation to orbital phase. No companions were detected out to 200~mas. We used the Absil \citep{absil} and  Injection \citep{candid} methods to determine detection limits. The $5\sigma$ detection limit at 200~mas ranged from $\Delta m \approx 2.2$~mag for the 2022~March epoch (the poorest quality data night, with a single calibrator and long gap between target and calibrator) to $\Delta m \approx 5.8$~mag for the 2024 June 22 observation, with typical limits of $\Delta m \approx 4$--5~mag.

\subsection{Angular Diameters}

\begin{deluxetable*}{l l l l l l l l l l}
\tablecaption{Measured limb darkened (linear law) disk angular diameters. \label{tab:diameters_ld}}
\tablewidth{0pt}
\tablehead{
\colhead{Date} & 
\colhead{Phase\tablenotemark{a}} & 
\colhead{$\theta_{\text{LD,H}}~(\text{mas})$} & 
\colhead{$V^{2}_{0,\text{H}}$\tablenotemark{b}}  &
\colhead{$\chi^{2}_{\text{H}}$} & 
\colhead{$\theta_{\text{LD,K}}~(\text{mas})$} &  
\colhead{$V^{2}_{0,\text{K}}$\tablenotemark{b}} &
\colhead{$\chi^{2}_{\text{K}}$} & 
\colhead{$R_{\text{H}}~(R_{\odot})$\tablenotemark{c}} & 
\colhead{$R_{\text{K}}~(R_{\odot})$\tablenotemark{c}} 
}
\startdata
2022-03-03 & 0.77 & $0.71 \pm 0.06$ & 0.87 & 3.45 & ---             & ---  & ---  & $70 \pm 6$ & ---        \\
2023-05-09 & 0.67 & $0.74 \pm 0.01$ & 0.95 & 1.51 & $0.81 \pm 0.04$ & 0.96 & 1.54 & $72 \pm 1$ & $80 \pm 4$ \\
2023-06-04 & 0.78 & $0.73 \pm 0.04$ & 1.10 & 6.44 & $0.77 \pm 0.02$ & 1.01 & 1.57 & $72 \pm 4$ & $76 \pm 2$ \\
2023-09-09 & 0.21 & $0.72 \pm 0.01$ & 0.82 & 1.89 & ---             & ---  & ---  & $71 \pm 1$ & ---        \\
2024-06-22 & 0.47 & $0.64 \pm 0.01$ & 1.04 & 0.70 & $0.66 \pm 0.01$ & 0.98 & 0.77 & $63 \pm 1$ & $65 \pm 1$ \\
2024-06-23 & 0.47 & $0.68 \pm 0.01$ & 1.07 & 1.85 & $0.67 \pm 0.02$ & 1.05 & 1.24 & $66 \pm 1$ & $66 \pm 2$ \\
2025-02-26 & 0.56 & ---             & ---  & ---  & $0.67 \pm 0.02$ & 0.87 & 2.85 & ---        & $66 \pm 2$ \\
\enddata
\tablenotetext{a}{Orbital phase calculated using $T_{0}~(\mathrm{HJD}) = 2459978.37$ \citep{2025munari}, adding 0.25 so that $\phi=0$ is at superior conjunction (RG behind WD).}
\tablenotetext{b}{Fitted zero-baseline visibility; deviations from unity reflect calibration systematics.}
\tablenotetext{c}{Radii are calculated using $d=914^{+24}_{-22}$ pc \citep{schaeferdistances}.}
\end{deluxetable*}
We measured angular diameters using the Parametric Modeling of Optical InteRferomEtric Data module (PMOIRED; \citealp{antoine_merand_2024_10889235})\footnote{\url{https://github.com/amerand/PMOIRED}}, which uses gradient descent for $\chi^{2}$ minimization and determines uncertainties with bootstrapping. PMOIRED offers a variety of models, such as uniform disks, user specified intensity profiles, ellipsoids, and rings and also allows for the joint fitting of models. Our data contain only the first lobe of the  $V^2$ curve. In stars, the second and later lobes provide information on scales smaller than the stellar disk. This, in combination with nearly zero closure phases means that the observations do not contain enough information to constrain models such as disks or triaxial ellipses. Indeed, when we fitted the data to ellipsoidal models, the resulting values for the parameters were highly unconstrained, with large uncertainty. This does not mean the data rule out a non-spherical and non-symmetric", only that we could not constrain any deviation from centro-symmetry at the angular resolution of CHARA given the angular size of the star.

Likewise, limb darkening coefficients, when left as free parameters, were unconstrained because the second lobe  of the $V^2$ curve provides important information on stellar limb darkening. We were unable to fit limb darkening coefficients for multiple laws, including the \citet{hestlaw} power law, when coefficients were left as free parameters. Therefore, we ran fits using the linear, quadratic, square root, logarithmic, and four parameter laws described in \citet{claret2011}, fixing the limb darkening coefficients to those provided in the database that accompanies that paper \citep{claret2011database}.  In these fits, we used the coefficients for $T_{\text{eff}}= 3500-3750$ K,  $\log g$=0.5-1.0 , Z=0.2 stars, based on the RG's stellar parameters in \citet{2025hinkle}. We found that the limb darkened disk (LDD) angular diameters did not vary by more than $1\sigma$ regardless of the law used. Similarly, when we translated the linear coefficient into a fixed $\alpha$ for use in the \citet{hestlaw} power law, we found angular diameters within $1\sigma$ of fits made using other laws.

Because we have limited spatial coverage and the different limb darkening profiles resulted in angular diameters within $1\sigma$ of each other, here we report the angular diameters resulting from using the linear limb darkening law for the $T_{\text{eff}}= 3500$ K, $\log g$=0.5, Z=0.2 model using \textit{H}-band coefficient $u=0.4827$ and \textit{K}-band coefficient $u=0.3871$ because this is the simplest law. In addition to angular diameter, we left the zero-baseline visibility squared parameter $V^2_0$ free in order to absorb any residual transfer function errors due to calibration systematics. In all fits, we used 200 bootstrapping iterations to determine uncertainties in the angular diameters and incorporated a 5\% error floor for $V^{2}$. We did not use closure phases in the fit because they would not add additional information for fitting the angular diameter of the star at this angular resolution. After fitting angular diameters, we followed the instructions of the MIRC-X-MYSTIC Pipeline User Manual and divided the \textit{H}-band angular diameters by $1.0054\pm 0.0006$ for \citep{2022gardner} and the \textit{K}-band angular diameters by $1.0067\pm0.0007$ to account for wavelength calibration.  

The average LDD angular diameter we measure for the RG in T CrB is $0.70\pm0.04$ mas in \textit{H}-band and $0.72\pm0.07$ mas in \textit{K}-band across all epochs and $0.72\pm0.01$ mas in \textit{H}-band and $0.79\pm0.03$ mas in \textit{K}-band excluding the 2024-2025 data, which used calibrators close in angular diameter to the RG. In this paper, we adopt a distance $d=914^{+24}_{-22}~\textrm{pc}$ that \citet{schaeferdistances} determined using the \textit{Gaia} DR3 parallax, a galactic disc prior, and a prior based on the peak magnitude of the 1946 eruption. Using this distance, the corresponding physical radii are $69\pm5~R_{\odot}$ in $H$-band and $71\pm8~R_{\odot}$ in $K$-band across all epochs and $71\pm3~R_{\odot}$ in $H$-band and $78\pm4~R_{\odot}$ in $K$-band excluding the 2024-2025 data. Note that  \citet{2025munari} used $d=920$ pc. \citet{2025hinkle} used a distance from \citet{bailer-jones_estimating_2021} ($887^{+22}_{-23}$ pc), which was derived using a galactic prior and is consistent with the distance derived using the \textit{Gaia} DR3 parallax after applying a zero-point correction as described in \citet{2021A&A...649A...4L} ($\varpi_\mathrm{corr} = 1.129 \pm 0.028$ mas,  $d \approx 885$ pc). Using any of these other distances changes our derived radius by less than 4\%, within our measurement uncertainty.

\citet{2025hinkle} adopted 
$887^{+22}_{-23}$ pc from \citet{bailer-jones_estimating_2021}, \textit{Gaia} DR3 parallax zero-point correction, which yields 
$\varpi_\mathrm{corr} = 1.129 \pm 0.028$ mas ($d \approx 885$ pc; 
\citealt{2021A&A...649A...4L}).  Using any of these distances changes our derived radius 
by less than 3\%, within our measurement uncertainty.

All these distances rely primarily on the \textit{Gaia} DR3 single-star astrometric solution, which is likely biased by the binary's photocenter motion given the system's high re-normalized unit weight error (RUWE=1.64) and a parallax comparable to the semi-major axis ($\varpi=1.09\pm0.03$ mas vs $a=1.05\pm0.03$ mas using $a=206.4~R_{\odot}$from \citet{2025munari}). As noted by \citet{2025hinkle}, an astrometric binary solution in a future \textit{Gaia} data release will provide a more reliable measurement of the distance.

In Table \ref{tab:diameters_ld}, we present the measured angular diameters as well as these fit $V^{2}_0$. In Figure \ref{fig:diamvsphase}, we present a plot of the LDD angular diameters against phase and date of observation. The phases in the table and figure have been calculated using the orbital parameters of \citet{2025munari}, but the phase was adjusted by 0.25 so that 0 corresponds to superior conjunction, when the RG is behind the WD. 

The \textit{K}-band angular diameters are on average 8\% larger than those in \textit{H}-band for fits to the 2022 and 2023 observations. In practice, the use of different limb darkening coefficients should account for wavelength dependent variations. To test this, we also ran fits to obtain a single diameter across both bands for nights on which both MIRC-X and MYSTIC were used. In these fits, we matched limb darkening coefficients to their corresponding wavelength ranges (i.e. we still used different coefficients for \textit{H}-band or \textit{K} band), running fits for both fixed (using the previously described \citet{claret2011} values) and free coefficients, with the stipulation that the fit should give the same diameter for both \textit{H} and \textit{K}-band data. For each set, we ran with a fixed $V^2_0=1$ and a free $V^2_{0}$ using different values for each instrument. We found that forcing a single LDD to simultaneously describe both bands degraded the fit quality for nearly every epoch, regardless of the coefficients used. The angular diameters measured when using this approach fell between those obtained when fitting each band separately.

Although the \textit{H}-band contains mainly continuum portions of the spectrum for M3III stars like the RG, the \textit{K}-band includes CO bands in these stars, including T CrB \citep{2019MNRAS.486.3498E}. In order to test whether the CO first-overtone bandheads ($\lambda \approx 2.29$--$2.39\,\mu$m) were the cause of larger angular diameters in the \textit{K}-band data, we compared angular diameters measured within only this region and those obtained excluding it. The angular diameters measured on data that excluded the CO bandheads did not change significantly from those obtained using the full \textit{K}-band data. Angular diameters measured using data that included only the wavelengths covering the CO bandheads were only slightly larger than those derived using the full \textit{K}-band or CO-excluded portions (e.g. 0.01 mas, $<2\%$) apart from that measured using data collected in 2023-05-09, which had an angular diameter 0.08 mas ($\sim10\%$) larger. However, this fit was made using only 88 $V^{2}$ measurements. Such a sizable difference between CO and continuum \textit{K}-band angular diameters was not found for the observations made on 2023-06-04, which had 232 $V^{2}$ points in the CO region and was only one month later, so the larger angular diameter in the CO band in 2023-05-09 data is likely an artifact of fitting to sparse data rather than an observation of molecular extension.

\subsection{Simulations}\label{sec:sims}

To test our fitting technique on known shapes, we used ROTIR\footnote{https://github.com/fabienbaron/ROTIR.jl} \citep{Martinez2021} and OITOOLS\footnote{\url{https://fabienbaron.github.io/OITOOLS.jl/dev/}} \citep{Martinez2021} to generate synthetic observations of Roche lobe filling stars. The simulation grid consisted of models spanning inclination from face-on $i=50^{\circ}-65^{\circ}$, position angle (degrees East of North) PA$=0^{\circ}-165^{\circ}$, $d=752-960$ pc, and $M_{\text WD}=1.25-1.35 M_{\odot}$ with mass ratio $q$ and semi-major axis $a$ derived from the mass function $f(m)=0.322 M_{\odot}$\citep{2025munari}. For other binary parameters we used values from \citet{2025munari}, including eccentricity $e=0$, period $P=227.5538$ days, and $T_{0}=2459978.37$ HJD. Note that ROTIR uses the same convention for $T_{0}$ as \citet{2025munari}, time of passage of ascending quadrature. We used the (\textit{u},\textit{v}) coverage and error statistics of the actual observations to generate the synthetic data for all nights on which MIRC-X was used, thereby producing simulations at different orbital phases. In order to test the impact of (\textit{u},\textit{v}) coverage on the measured angular diameters, we also generated a smaller set of simulations simulating 4 hours of observing time, which is longer than we were able to obtain on each night of observing. The models were generated for 60 different equally-spaced phases across the full orbit and used the parameters of \citet{2025munari} ($i=61.5^{\circ}$,$M_{\text WD}=1.35 M_{\odot}$) at four PAs ($0^{\circ}, 45^{\circ},90^{\circ},135^{\circ}$). We also generated a control simulation of a LDD sphere with $\theta_{\rm LD} =  0.73$~mas, which corresponds to an angular diameter equal to twice the Roche volume radius calculated using the approximation of \citet{1983eggleton} with $q = M_{\rm RG}/M_{\rm WD}=0.69$ and $a=206.4~R_{\odot}$ at $d = 914$~pc. All simulations were fit with the same procedure as the data except with $V^{2}_0$ fixed  because the simulations did not carry over calibration errors from the data.

 Fits of an LDD to the control simulation of a sphere resulted in a mean $\theta_{\text {LD}}=0.73\pm0.01$ mas, which matches the input angular diameter. However, Roche lobe filling stars are not centro-symmetric, so fitting an LDD model introduces a projection-dependent bias. The measured LDD will depend on the inclination, PA, and orientation of the projected object relative to the baseline coverage at each observed epoch. We found that for models spanning $i \approx 57^\circ$--$65^\circ$ and $d \approx 890$--$960$~pc (which covers parameters reported in \citealt{2025munari, 2025hinkle, schaeferdistances}), the fitted LDD ranges from 0.687--0.769 mas at the phases corresponding to the 2022-2023 observations and 0.627--0.755 mas for the phases corresponding to the 2024 observations. For the parameters of \citet{2025munari} ($i=61.5^\circ$, $d=914$~pc, $M_{\rm WD}=1.35~M_\odot$), the range is 0.709--0.771~mas across the sampled PAs for the 2022--2023 phases and 0.690--0.706 mas for the 2024 phases. 
  
Fits to the simulations using a larger (\textit{u},\textit{v}) coverage produced LDDs that differed from those fit simulations based on our (\textit{u},\textit{v}) coverage by $-9\%$ to $+8\%$ depending on PA, with the largest differences at PA$= 135^\circ$ and $45^\circ$ respectively. This difference reflects the predominantly east-west orientation of the baselines in the observations. This means that although the angular diameters in this letter measure the angular size of the RG as sampled by the available baselines, a more complete coverage could shift the inferred diameter by up to $\sim$$9\%$ depending on the PA of the system. 

Nonetheless, the angular diameters (0.712--0.734~mas) measured with 2022-2023 fall within the range predicted by Roche lobe filling simulations using the observed $(u,v)$ coverage (0.687--0.769~mas) and within the range predicted by two of the four position angles sampled in the synthesized-track simulations (PA $= 0^\circ$ and $90^\circ$), supporting a Roche lobe RG. Some models also replicated both the ranges of the angular diameters measured with 2022--2023 data as well as the 2024 data. These include those with lower inclinations or lower WD masses but because we are skeptical of the calibration on those data, we caution against interpreting this as an indication that the data favor these system parameters.

\section{Discussion} \label{sec:discussion}
\begin{figure*}[!h]
    \gridline{
        \fig{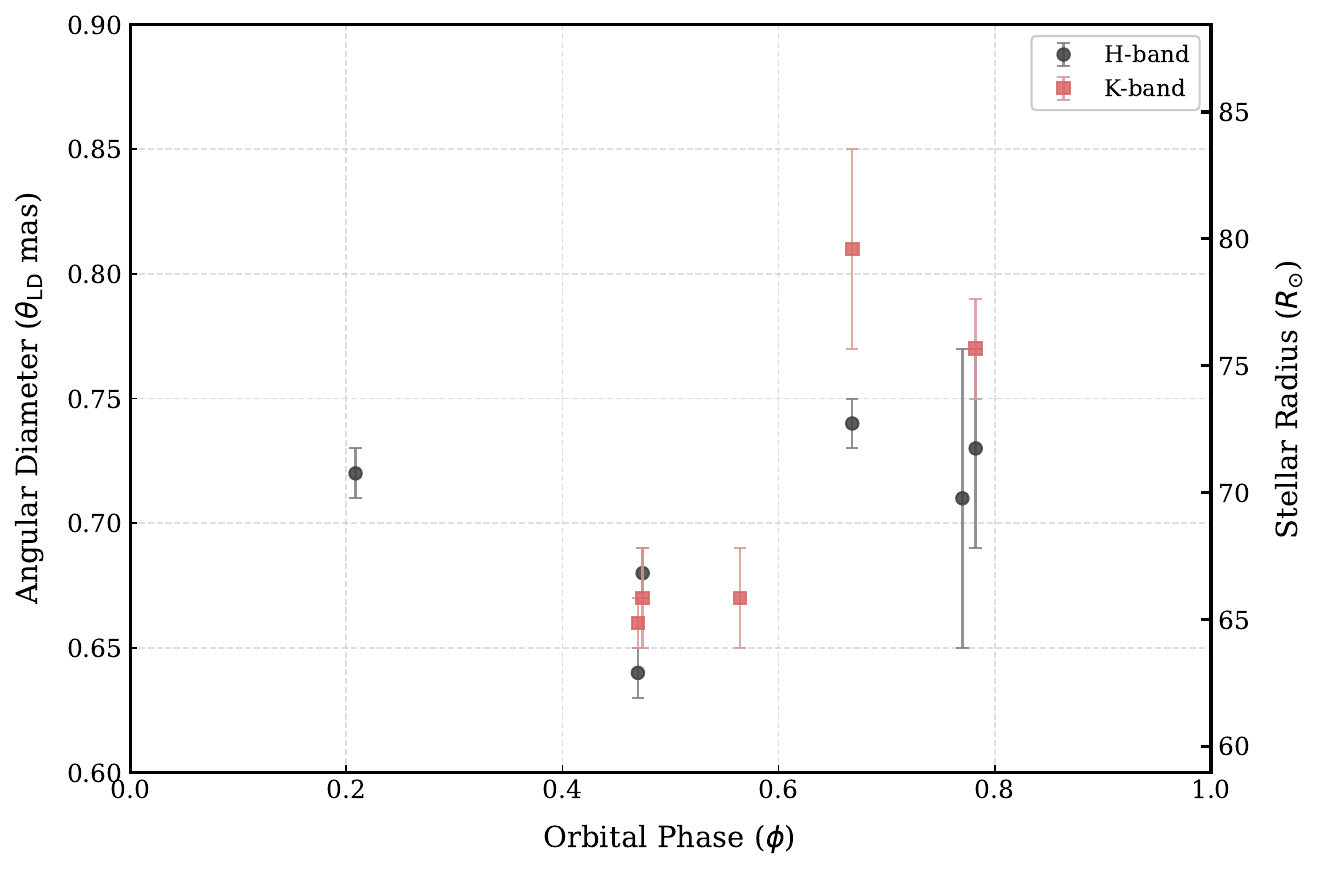}{0.75\textwidth}{(a) Variation in angular diameter with phase.}}
        \gridline{
        \fig{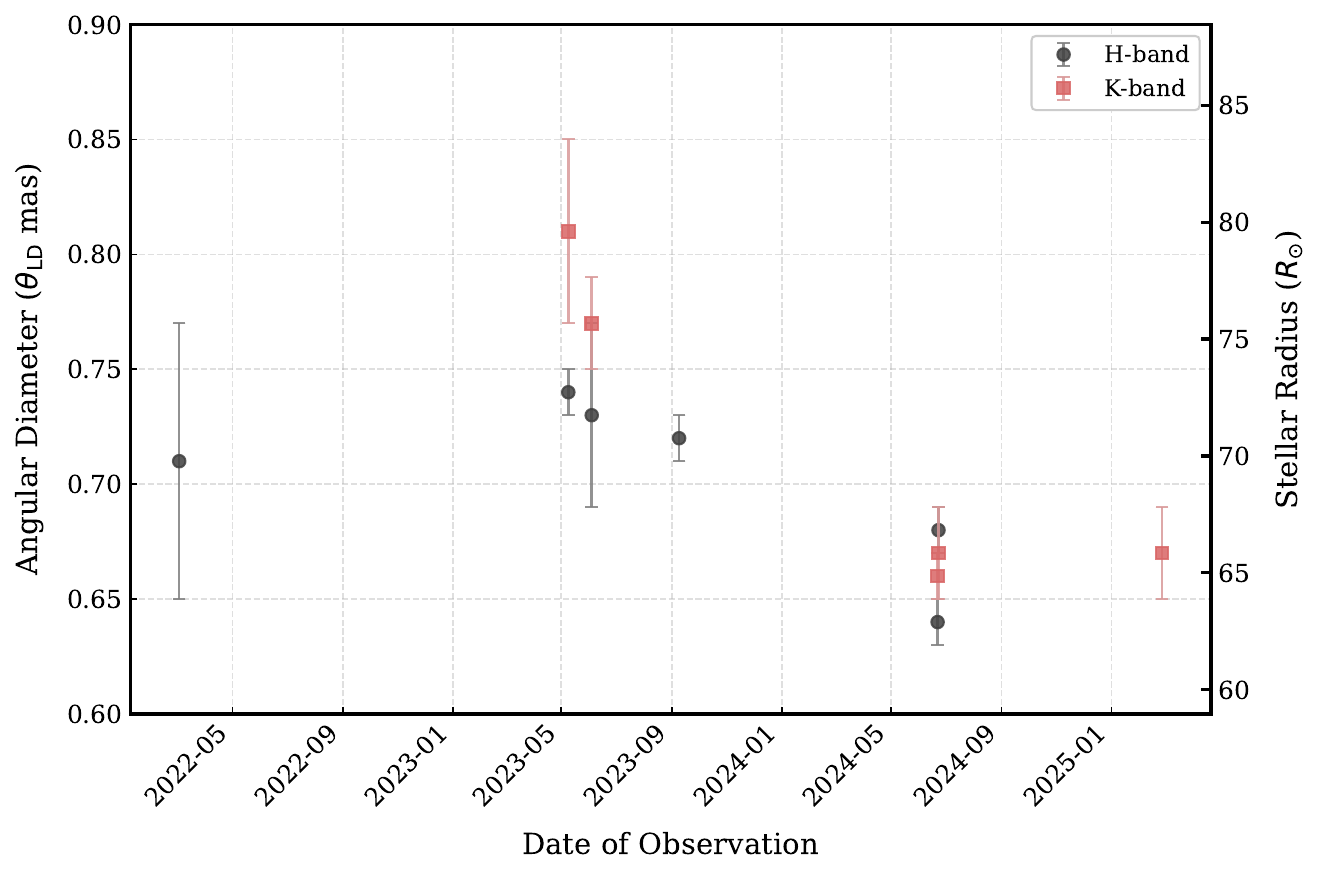}{0.75\textwidth}{(b) Variation in angular diameter with date.}
    }
    \caption{Variation in angular diameter with phase and date of the LDD angular diameters fit to the observations reported in this paper.}
    \label{fig:diamvsphase}
\end{figure*}
The phase-based variation in angular diameter presented in Table \ref{tab:diameters_ld} and Figure \ref{fig:diamvsphase} seems to follow what might roughly be expected for an elongated star. The largest angular diameters are measured closer to quadrature phases and the smallest close to inferior conjunction. This follows a similar trend (brighter near quadrature, dimmer near conjunctions) found in \textit{H} and \textit{K}-band photometric data of the RG \citep{maslennikova_recurrent_2023}. However, the observations taken close to inferior conjunction used calibrator stars with angular diameters close to that of T CrB, which makes the angular diameters obtained from these data less reliable and more sensitive to diameters used in the calibration process. The smaller diameters measured in 2024 and 2025 do fit within or close to the ranges found by fits to simulations at some PAs (0.627--0.755~mas for system parameters consistent with the literature, and 0.690--0.706~mas for the parameters of \citealt{2025munari}), but the unreliable calibration prevents a firm conclusion. 

Light curves \citep[e.g.,][]{shahbaz,maslennikova_recurrent_2023,2025hinkle} support the presence of a Roche-filling RG in T CrB. In addition, \citet{2025munari} used a combination of \textit{UBVRI} photometry, optical spectroscopy, and ultraviolet observations along with radiative modeling to describe the system during its quiescent, ``super-active accretion phase" (SAP, May 2015 to April 2023), ``deep minimum" (May 2023 to late 2024), and recovery state (early 2025), finding support for a Roche-filling RG of $R= 71~R_{\odot}$. The radii we derive from the observations reported here agree with this value, particularly using the \textit{H}-band observations ($R=69\pm5~R_{\odot}$), which mainly span the continuum of the spectrum of the RG. Further evidence for a Roche-filling RG comes from the simulations detailed in Section \ref{sec:sims}. These models show that the RG's angular diameters align with those expected for a Roche-filling star across a range of PAs and system parameters.

The angular diameters reported here come from measurements during three different periods of activity (the SAP, ``deep minimum,'' and subsequent recovery). There is no evidence in the literature to suggest that the radius of the RG changed during these stages, so we take our measurements to be representative of the typical radius of the star. The observational changes in the system during the past 11 years were induced mostly by behavior in the disk, such as the inside-out collapse of the disk that took place during the SAP \citep{2025munari}. These events were quite apparent at X-Ray, UV, and \textit{UBV} wavelengths \citep[e.g.,][]{2026A&A...707A.102L}, but at longer wavelengths, the system appeared rather stable throughout these stages. For example, \textit{H} and \textit{K}-band photometry reported in \citet{maslennikova_recurrent_2023} shows no statistically significant deviation in mean brightness or ellipsoidal modulation before, during, or after the SAP \citep[see also][for a comparison of the \textit{J} and \textit{K}-band phototometry in that paper to an ellipsoidal light curve, showing residuals close to zero]{2025hinkle}. Moreover, using radiative modeling, \citet{2025munari} found that the small increase in \textit{I}-band magnitudes during the SAP could be explained by a contribution from the accretion disk to the lightcurve of the RG, rather than changes in the RG. There is evidence of an irradiated patch on the RG on the side facing the disk, which reached a higher temperature during the SAP \citep{2025planquart}, but in the model of \citet{2025munari}, this changed the surface-averaged temperature of the RG by $\sim60$ K, which is too small to measurably affect the angular diameter. Although \citet{2025planquart} found a strengthening of \ion{O}{1} triplet (7772 \AA) absorption on the giant's irradiated hemisphere during the SAP, they did not find evidence for a change in log $g$ or surface integrated $T_{\text{eff}}$, which might suggest a change in radius. Instead, they show that this feature resulted from an irradiated region with a temperature around 7500 K on the deformed side of the RG. Given its size and temperature, the photosphere of the RG ($T_{\text{eff}} \approx 3500\text{--}3600$,K; \citealp{2025munari, 2025hinkle}) would be significantly brighter than this feature in the \textit{H} and \textit{K} bands, so we do not expect it to have an impact on our measurements.
 
The \textit{K}-band diameters are on average 8\% larger than those in the \textit{H}-band. This may support the presence of atmospheric extension. For example, \citet{2019MNRAS.489.2595H} found uniform disk diameters to be 1-17\% larger in CO first overtone for $\gamma$ Cru, an M3.5 III star. However, measurements of the angular diameter of the RG in only the CO bandhead wavelengths did not find a significant difference in angular diameter for most epochs, so it is possible that this difference is related to an issue with the limb darkening coefficient used. Without more angular and spectral resolution, we cannot distinguish between the two possibilities.

The reported angular diameters are notable as there is only one other reported angular diameter for the RG in this system. \citet{2025hinkle} referred to the work of \citet{Rogge2011}, who reported an angular diameter of $0.85\pm0.15$ mas which was averaged across orbital phases and \textit{H} and \textit{K}-bands. This is larger than our \textit{H} and \textit{K}-band angular diameters at any phase. However, because the limiting angular resolution of PTI was 
$\theta \gtrsim 1.0$ \citep{Colavita_1999, 2008ApJS..176..276V}, an angular diameter of $0.85\pm0.15$ mas should be considered an upper limit.

\section{Conclusion} \label{sec:conclusion}
We measured angular diameters of the RG in T CrB using data obtained in 2022-2025, finding $\theta_{\text{LD,H}}$ ranging from 0.64--0.74 mas and $\theta_{\text{LD,K}}$ ranging from 0.66--0.81 mas. Although there is some uncertainty in the translation of angular diameters to a physical radius due to effects of PA on the projected radius of a Roche lobe filling star, the average \textit{H}-band radius of $69\pm5~R_{\odot}$ and  \textit{K}-band radius of $71\pm8~R_{\odot}$ support Roche lobe filling given the Roche lobe radius ($71~R_{\odot}$) found using the orbital parameters of \citet{2025munari}. Roche lobe filling is further supported by simulations, which show that the reported angular diameters fall within those obtained when fitting model Roche lobes filling stars at a variety of PAs. Because there is no evidence in the literature for a change in the diameter of the RG during the differing stages of activity in the past 11 years, these angular diameters should be considered representative of the typical state of the RG. These observations will provide a baseline for comparison when the system erupts.

While our observations provide information about the state of the RG before the eruption, they can also serve as a good foundation for a more in-depth analysis of the system once the outburst does occur. With measurements of the size of the RG, knowledge of the system's binary orbit, and the information gained from analyzing the size and layout of ejected material after the eruption, it will be possible to provide an accurate description of the system's post-outburst geometry. This is important, considering that material ejected during the outburst often exhibits a very complex form, significantly diverging from the simple model of a spherically symmetric explosion. Previous observations of novae have revealed that, in general, the morphology of their explosion can be described as a slower outflow focused in the orbital plane of the binary and a faster ejecta expanding perpendicularly to it, with binary nature of the system being considered as one of the main factors influencing this kind of geometry (see \citealt{2021ARA&A..59..391C} and references therein). However, this view is complicated in the case of symbiotic novae, where the expanding material interacts with the pre-existing wind of the RG, which in turn decelerates the ejecta and creates powerful shocks seen in radio and X-ray (e.g. \citealt{2006Natur.442..276S,2012ApJ...748...43N,2019MNRAS.490.3691D,2022ApJ...938...34O}). The impact of the RG on the morphology of the ejecta may therefore be significant and knowledge of its parameters will be helpful in later modeling of the evolution of geometry and kinematics of expanding material following the outburst.

\begin{acknowledgments}
We thank the anonymous referee for their helpful comments. RN acknowledges support from the National Science Foundation under Grant No. 2213518. MOH was supported by the Polish National Science Center grant 2019/32/C/ST9/00577 and the ``Initiative of Excellence - Research University” (ID-UB) program at A. Mickiewicz University in Poznań. AG acknowledges support from the Agencia Nacional de Investigaci\'on y Desarrollo (ANID) through FONDECYT Regular grant 1241073. JM was supported by the
Polish NCN grant 2023/48/Q/ST9/00138. JDM acknowledges funding for the development of MIRC-X (NASA-XRP NNX16AD43G, NSF-AST 2009489) and MYSTIC (NSF-ATI 1506540, NSF-AST 1909165). JLS acknowledges support from NASA grant 80NSSC26K0654. SK acknowledges funding for MIRC-X received funding from the European Research Council (ERC) under the European Union's Horizon 2020 research and innovation programme (Starting Grant No. 639889 and Consolidated Grant No. 101003096).  This work is based upon observations obtained with the Georgia State University Center for High Angular Resolution Astronomy Array at Mount Wilson Observatory.The CHARA Array is supported by the National Science Foundation under Grant No. AST-1636624 and AST-2034336.  Institutional support has been provided from the GSU College of Arts and Sciences and the GSU Office of the Vice President for Research and Economic Development. Time at the CHARA Array was granted through the NOIRLab community access program (NOIRLab PropID: 2022A-922062, PI: T. Gaudin; NOIRLab PropID: 2023A-216008, 2023B-722017, 2024A-791648, 2025A-330555, PI: R. Norris \& M. Otulakowska-Hypka).  This research has made use of the Jean-Marie Mariotti Center Aspro and SearchCal service. This work has made use of data from the European Space Agency (ESA) mission {\it Gaia} (\url{https://www.cosmos.esa.int/gaia}), processed by the {\it Gaia} Data Processing and Analysis Consortium (DPAC,
\url{https://www.cosmos.esa.int/web/gaia/dpac/consortium}). Funding for the DPAC
has been provided by national institutions, in particular the institutions participating in the {\it Gaia} Multilateral Agreement.  RN acknowledges the use of Claude Sonnet 4.6 (Anthropic), Opus 4.5 (Anthropic) and Gemini 3 Pro (Google) for assistance with development of Julia scripts, improvement of the format of plots,  table formatting, and proofreading. No part of the technical analysis or data interpretation was performed by these models.
;
\end{acknowledgments}

%

\vspace{5mm}
\facilities{CHARA (MIRC-X, MYSTIC)}

\software{OITOOLS, PMOIRED, ROTIR}



\bibliography{ref_new}{}
\bibliographystyle{aasjournal}

\appendix
\restartappendixnumbering 
\section{Additional Plots}\label{sec:all plots}
\begin{figure*}[!h]
    \centering
    \gridline{\fig{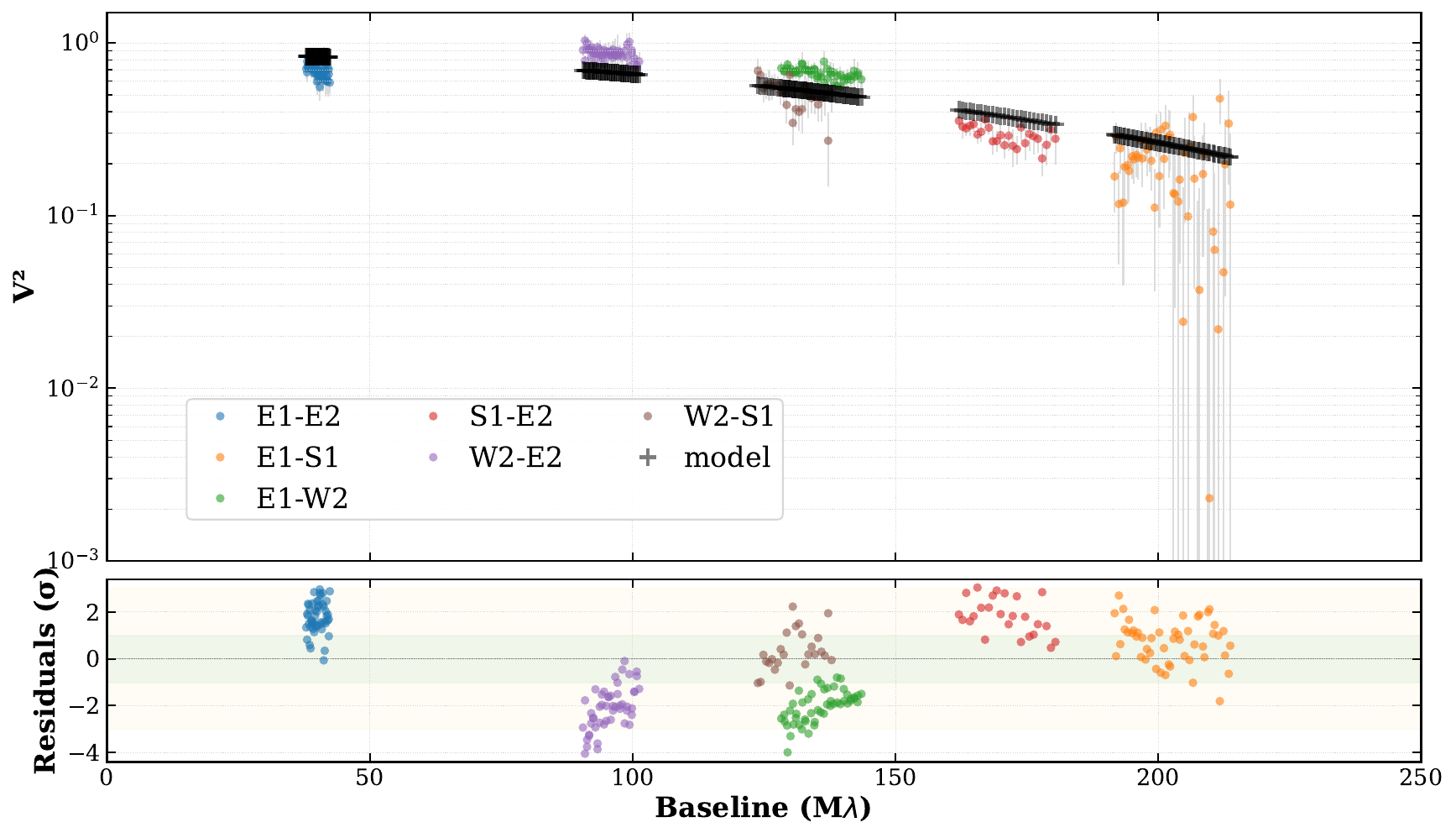}{0.45\textwidth}{(a) 2022-03-03 ($H$)}
              \fig{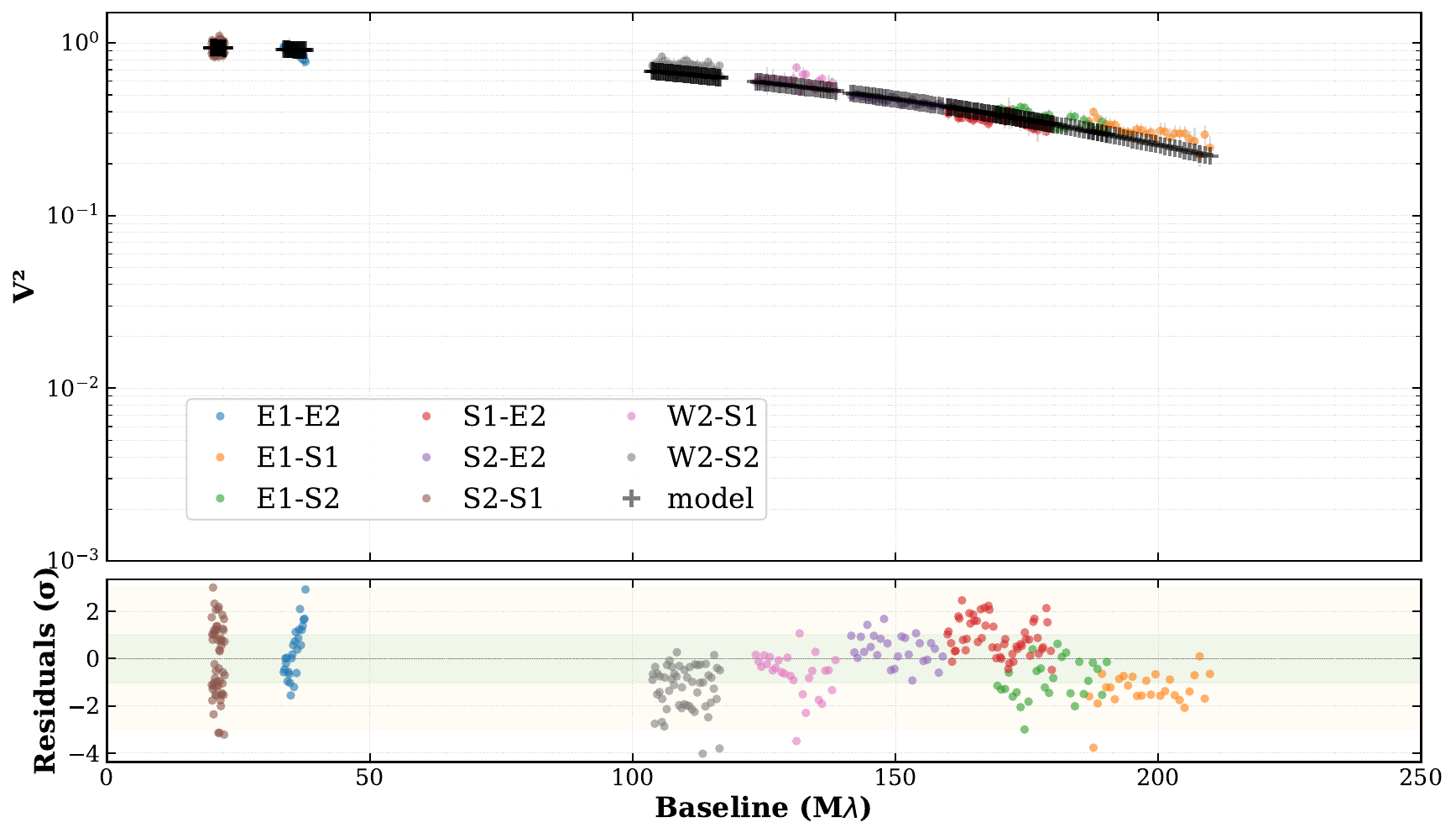}{0.45\textwidth}{(b) 2023-05-09 ($H$)}}
    \gridline{\fig{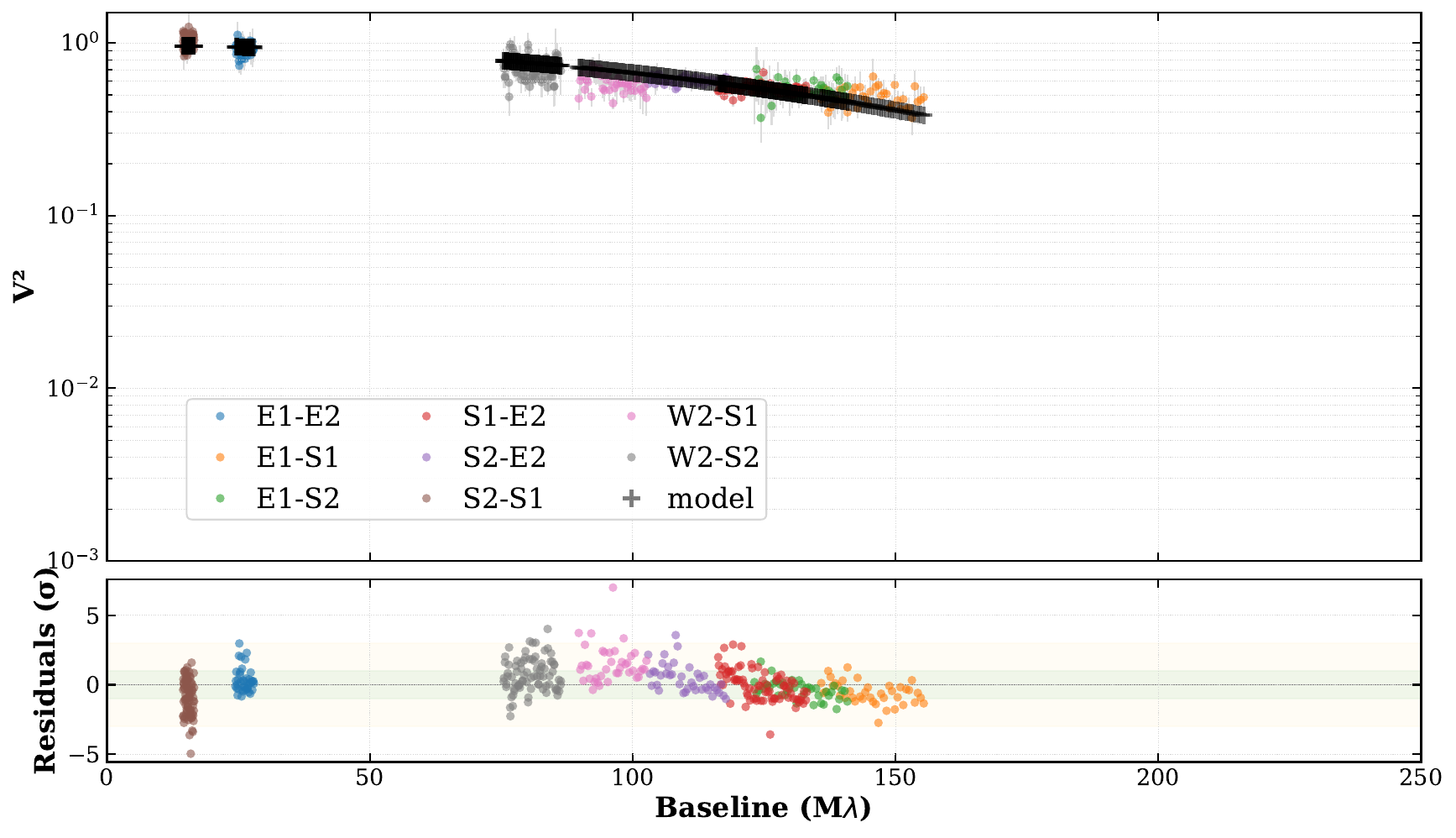}{0.45\textwidth}{(c) 2023-05-09 ($K$)}
              \fig{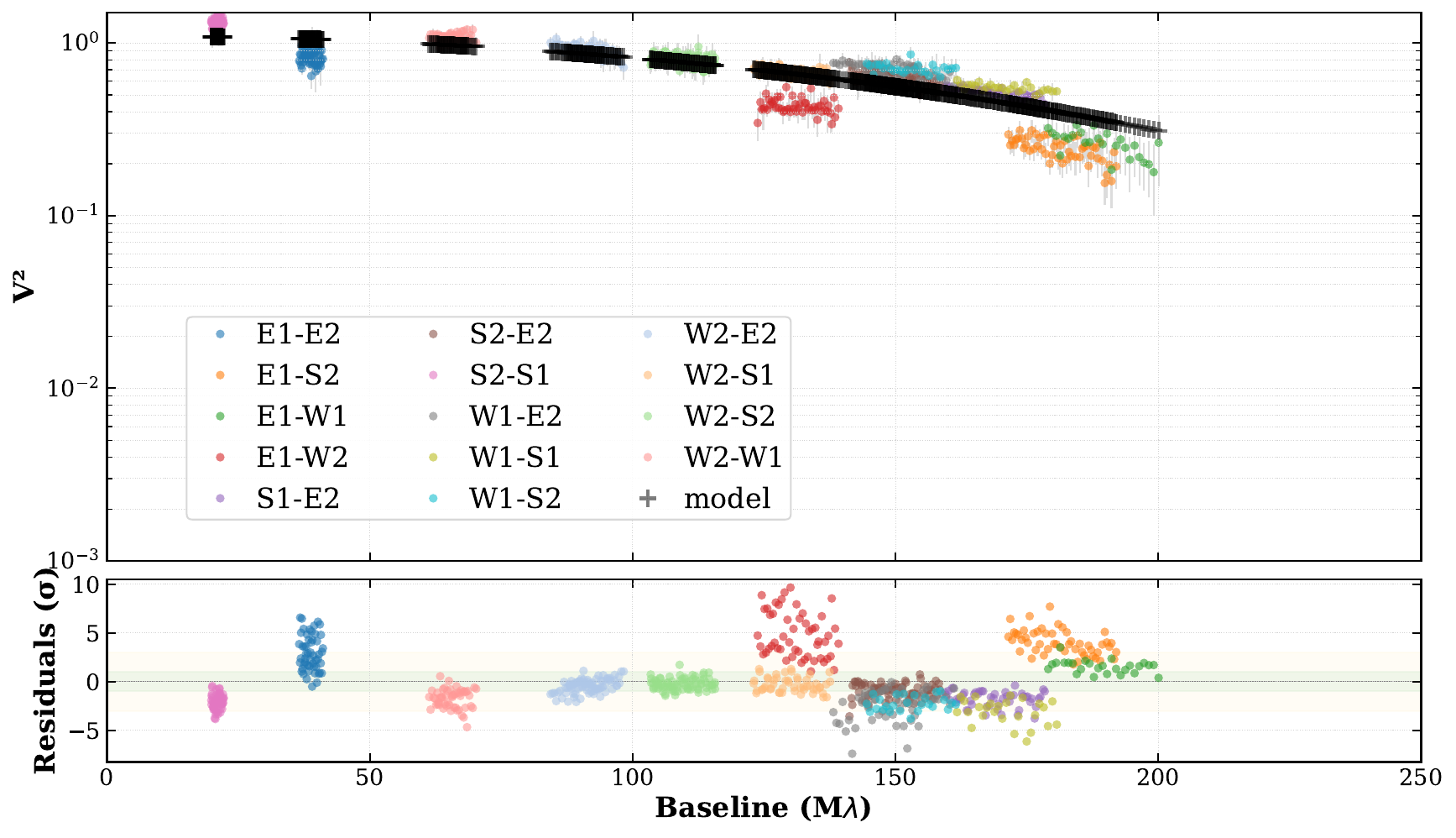}{0.45\textwidth}{(d) 2023-06-04 ($H$)}}
    \gridline{\fig{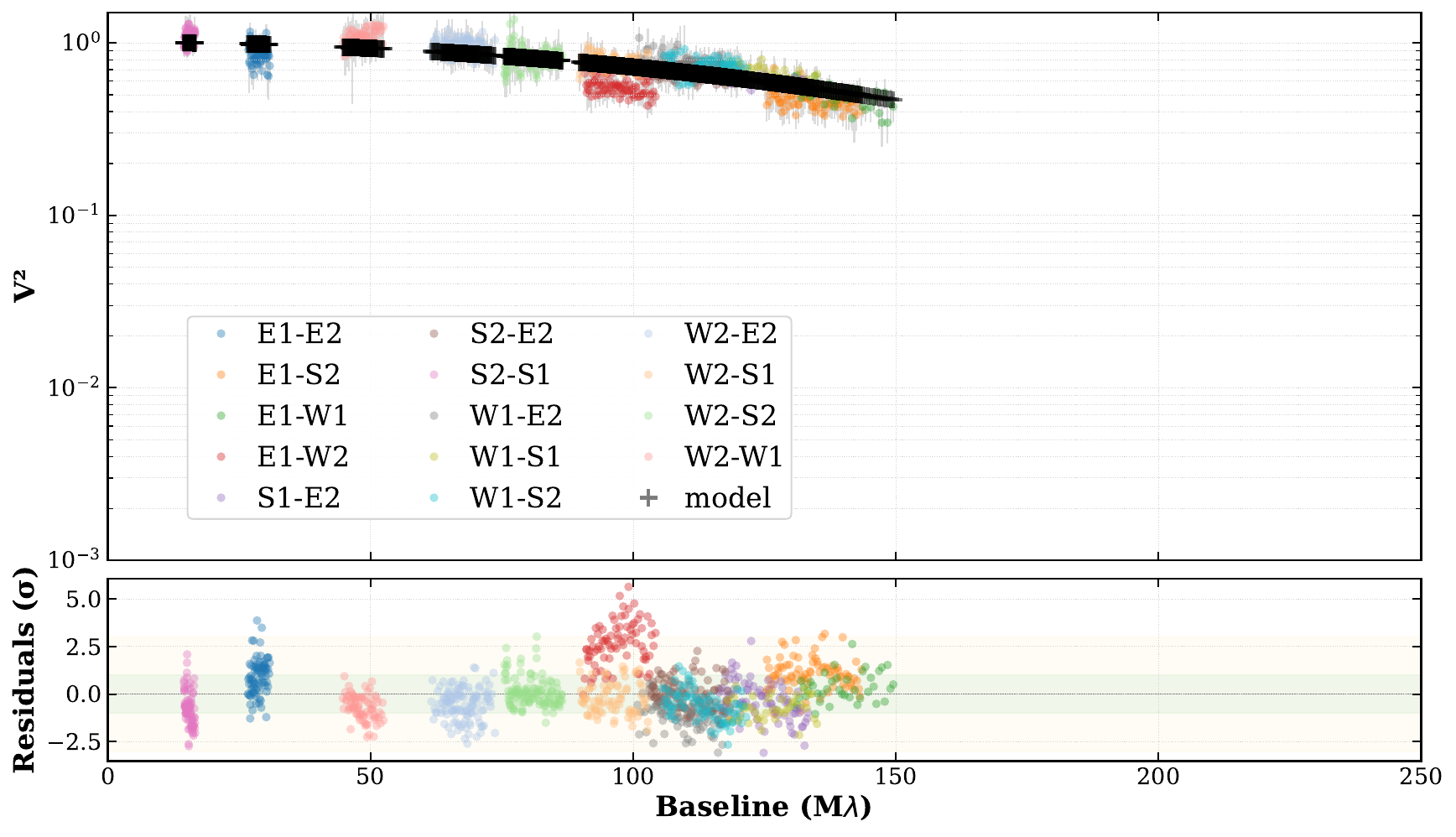}{0.45\textwidth}{(e) 2023-06-04 ($K$)}
              \fig{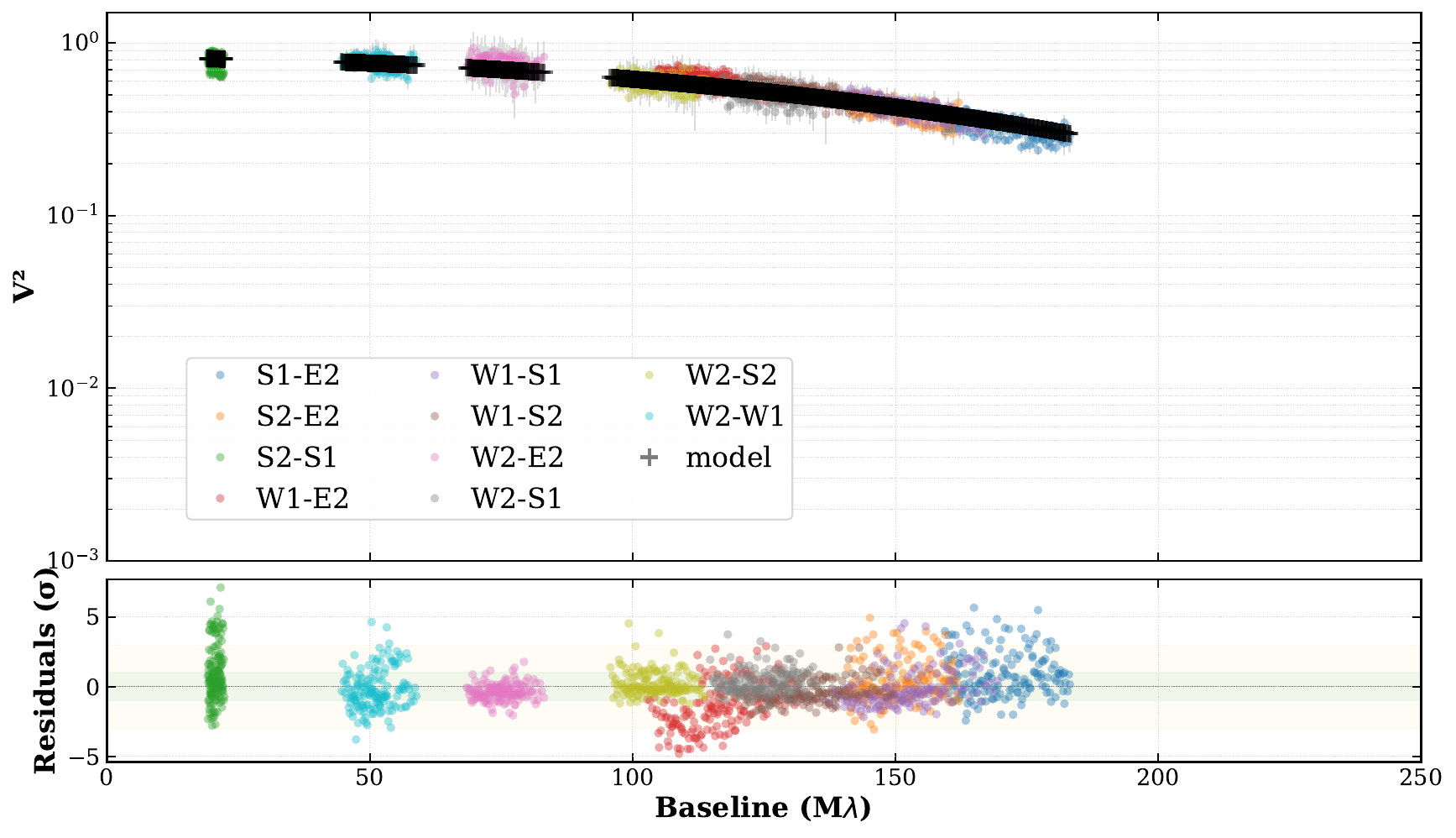}{0.45\textwidth}{(f) 2023-09-09 ($H$)}}
    \caption{Model fits and residuals for T\,CrB (part 1 of 2). The upper panel 
    of each subplot shows the squared visibility $V^2$ versus baseline length, 
    with data points colored by baseline pair and the best-fit limb-darkened disk 
    model (scaled by $V^2_0$) shown as black dots. The lower panel shows 
    residuals in units of $\sigma$; green shading indicates the $\pm1\sigma$ 
    region and yellow shading the $\pm3\sigma$ region.}
    \label{fig:tcrb_residuals_1}
\end{figure*}

\begin{figure*}[!p]
    \centering
    \gridline{\fig{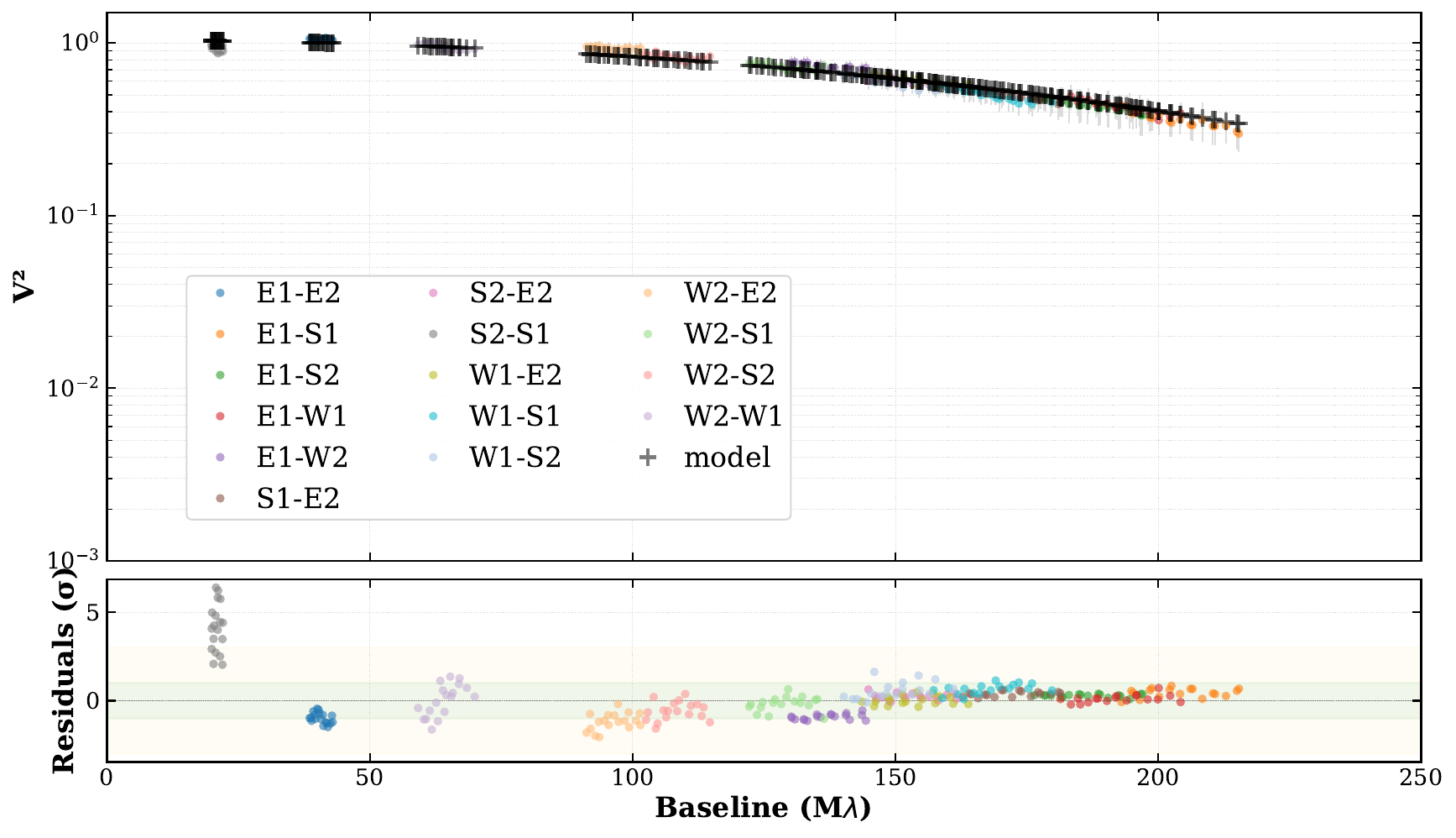}{0.45\textwidth}{(g) 2024-06-22 ($H$)}
              \fig{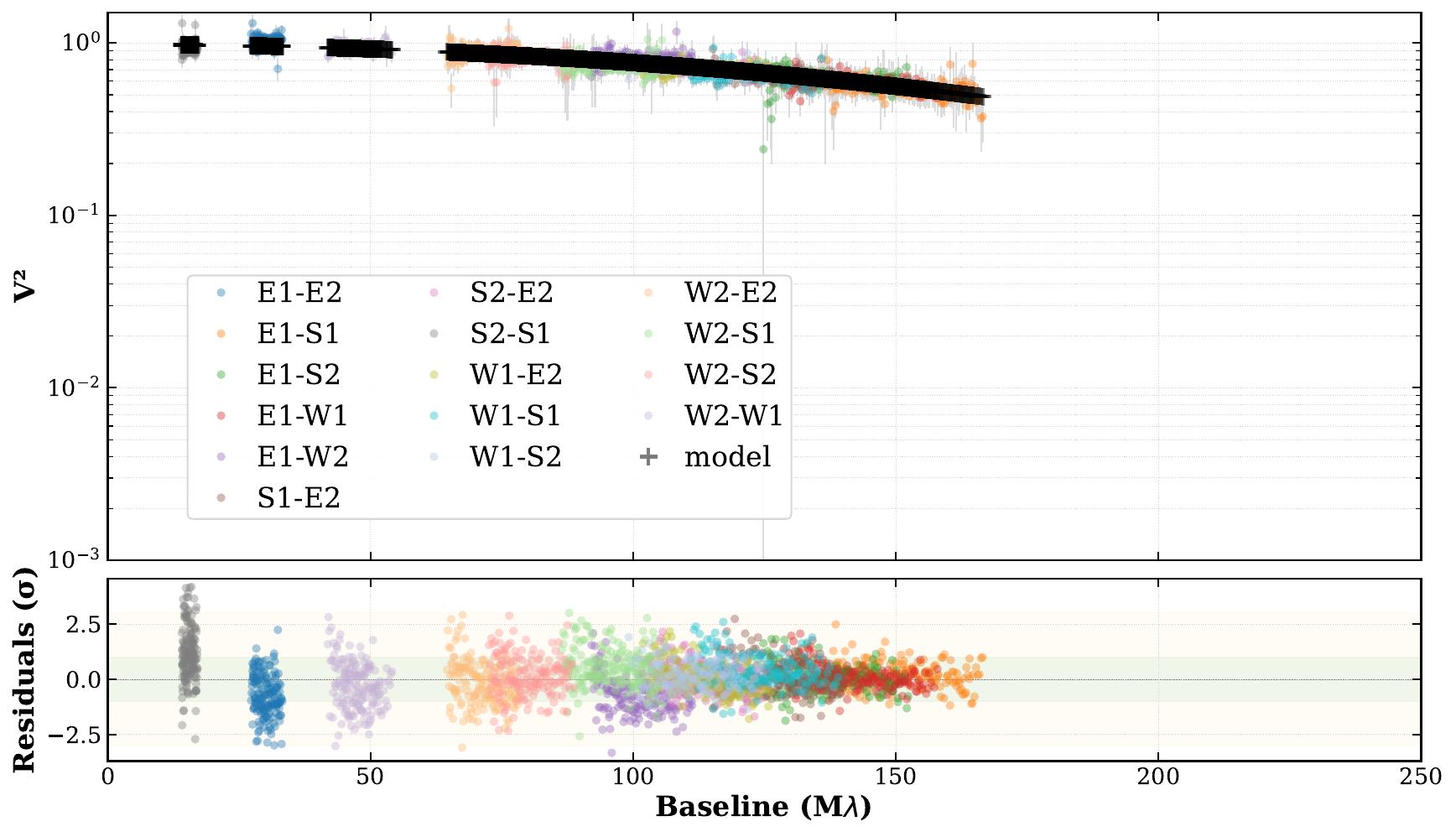}{0.45\textwidth}{(h) 2024-06-22 ($K$)}}
    \gridline{\fig{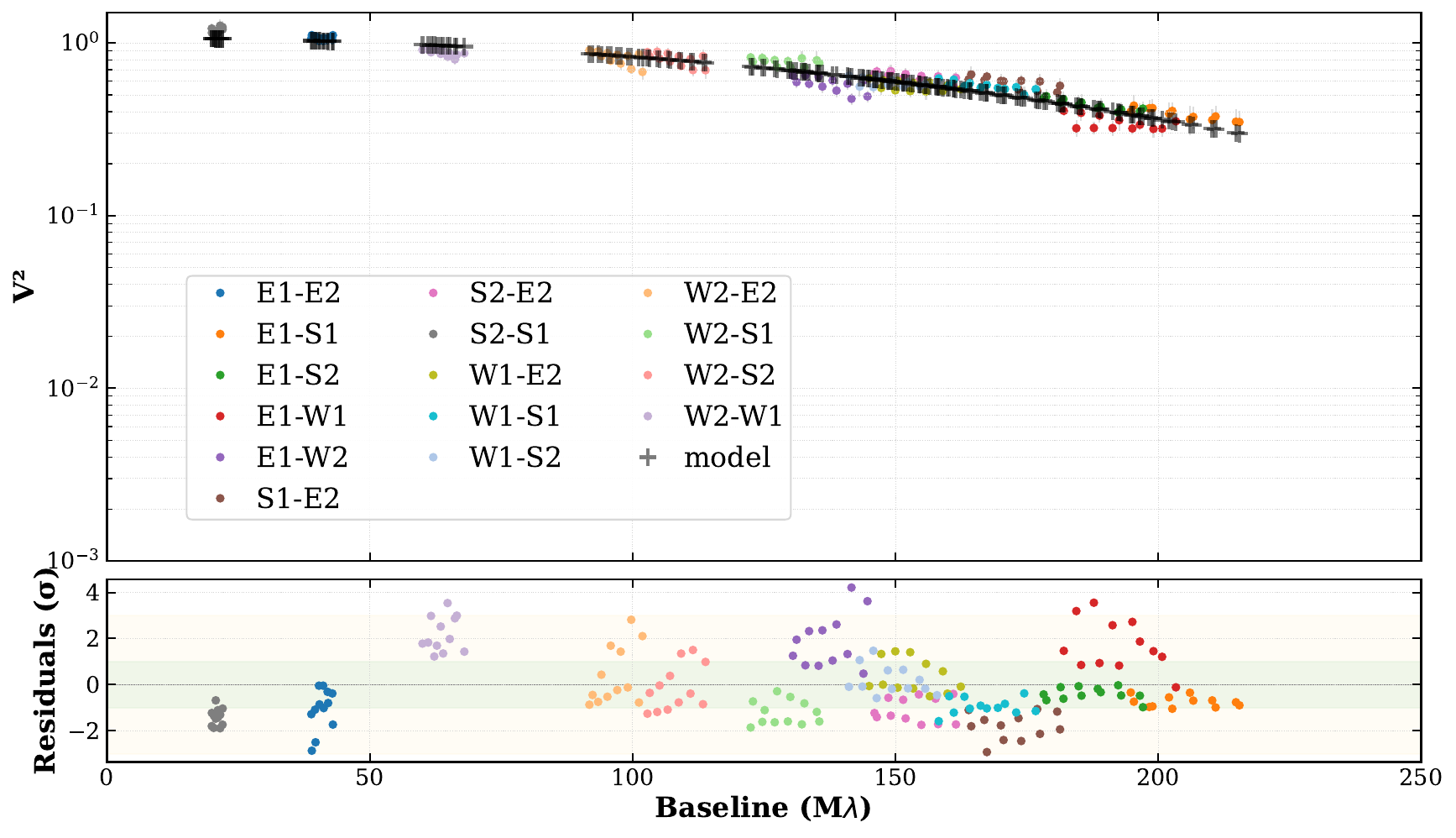}{0.45\textwidth}{(i) 2024 Jun 23 ($H$)}
              \fig{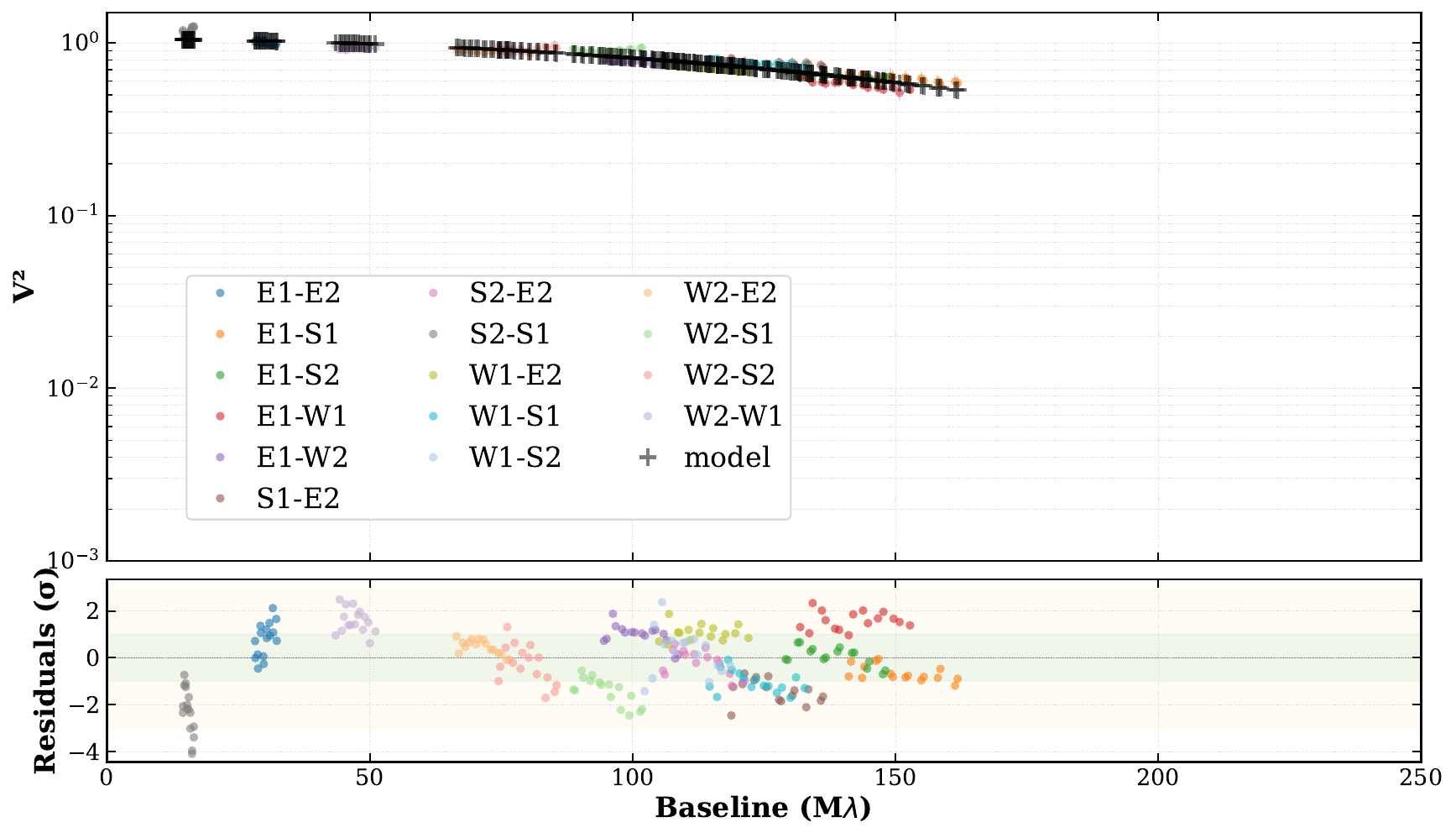}{0.45\textwidth}{(j) 2024-06-23 ($K$)}}
    \gridline{\fig{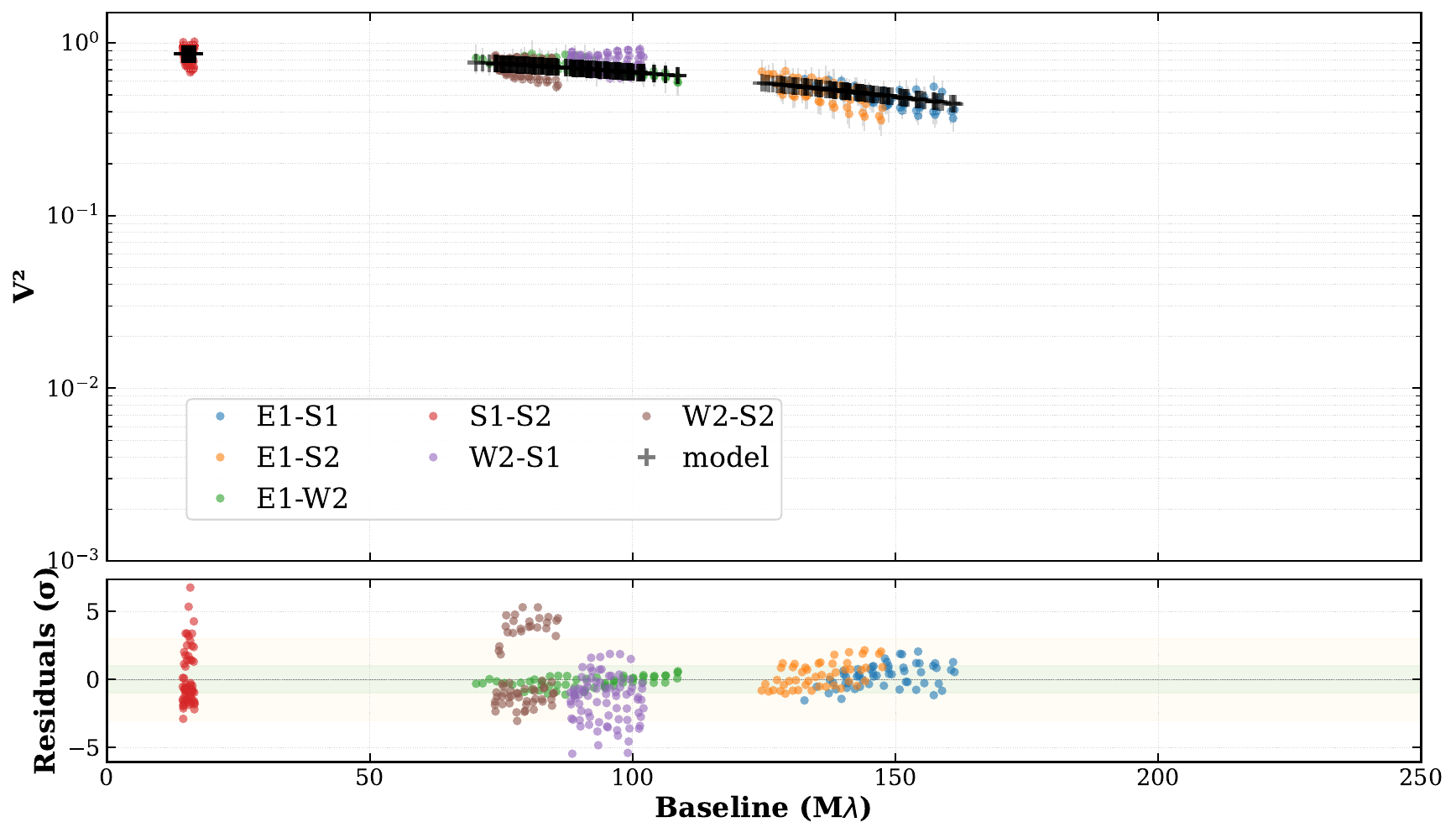}{0.45\textwidth}{(k) 2025-02-26 ($K$)}}
    \caption{Model fits and residuals for T\,CrB (part 2 of 2). See 
    Figure~\ref{fig:tcrb_residuals_1} for description.}
    \label{fig:tcrb_residuals_2}
\end{figure*}


\end{document}